\documentclass{article}
            \usepackage{graphicx}
            \usepackage{amsmath,amssymb}
\begin{document}
            \title{Fiber bundle description of number scaling in gauge theory and geometry.}
            \author{Paul Benioff,\\
            Physics Division, Argonne National
            Laboratory,\\ Argonne, IL 60439, USA \\
            e-mail:pbenioff@anl.gov}

            \maketitle

            \begin{abstract}
            This work uses fiber bundles as a framework to describe  some effects of number scaling on gauge theory and some geometric quantities. A  description of number scaling and  fiber bundles over a flat space time manifold, $M$, is followed by a description of gauge theory.  A fiber at point $x$ of  $M$  contains a pair of scaled complex number and vector space structures, $\bar{C}^{c}_{x}\times\bar{V}^{c}_{x},$  for each $c$ in $GL(1,C).$ A space time dependent scalar field, $g$, determines, for each $x,$ the scaling value of  the vector space structure that contains the value, $\psi(x),$ of a vector field at $x.$ Vertical components of connections between neighboring fibers are taken to be the gradient field,   $\vec{A}(x)+i\vec{B}(x),$ of $g.$ Abelian gauge theory for these fields  gives the result that  $\vec{B}$ is massless, and no mass restrictions for  $\vec{A}.$ Addition of an electromagnetic field does not change these results.  In the Mexican hat Higgs mechanism $\vec{B}$ combines with a Goldstone boson to create  massive vector bosons, the photon field, and the Higgs field.   For geometric quantities the fiber bundle is a tangent bundle with pairs, $\bar{R}^{r}_{x}\times\bar{T}^{r}_{x}$ for each $x$ and nonnegative real $r.$ $\vec{B}$ is zero everywhere. The $\vec{A}$ field affects path lengths and the proper times of clocks along paths. It also appears in the geodesic equation.  The lack of physical evidence for the  gradient $g$ field  means that either it couples very weakly to matter fields, or that it is close to zero for all $x$ in a local region of cosmological space and time.  It says nothing about the values outside the local region.

            \end{abstract}


            \section{Introduction}

            The use of fiber bundles \cite{Husemoller,Husemoller2} in the description of gauge theories and other areas of physics and geometry has grown in recent years \cite{Manin}-\cite{Pfeifer}.  In gauge theories connections between fibers in a bundle are described using Lie algebra representations of gauge groups. For Abelian theories the gauge group is $U(1)$ with $e^{i\alpha(x)}$ as a local gauge transformation.  For nonabelian theories \cite{Yang,Utiyama} the gauge group is $U(n)$ with the exponential of a sum over  generators of  $su(n)$ as the local gauge transformation. Additional details are given in many texts about field theories \cite{Cheng,Peskin,Montvay}.

            In this work the effect of number scaling on gauge theories and some geometric quantities is described. The base space, $M$, of the bundle is a flat manifold, either as Euclidean space or as  Minkowski space time. The  bundle fiber is expanded to include both scalar fields and vector spaces. This takes account of the fact that scalar fields are included in the description of vector spaces. For gauge theories a fiber at $x$  includes a complex scalar field, $\bar{C}_{x}$ with $\bar{V}_{x}.$ For tangent bundles each fiber  includes a real scalar field, $\bar{R}_{x}$  with a tangent space, $\bar{T}_{x}.$ This expansion is achieved by use of the fiber product \cite{Husemoller,Husemoller2} of two bundles. Because $M$ is flat  all the fiber bundles considered here are product bundles.

             Number scaling is accounted for by expansion of the bundle fiber to include  pairs of scaled scalar fields and vector spaces for all scaling factors. The scaling factors are complex for gauge theories and real for geometric properties.

            This work expands earlier work on number scaling by the author \cite{BenNS,BenNOVA,SPIE4} in that fiber bundles play an essential role. Scaling is accounted  for by use of a space or space time dependent real or complex valued scaling field. Vertical components of connections between fibers at neighbor locations on $M$ are based on this field.

            This work also differs from the earlier work in that the  vector scaling fields, as gradients of the scaling field are integrable.  In earlier work  vector scaling  fields were used with no connection to scalar fields.  As a result the integrability of the fields was open.

            Scaling is by no means new in physics and geometry. It  was used  almost 100 years ago by  Weyl \cite{Weyl}  to construct a complete differential geometry.\footnote{\label{EIN}Weyl introduced  a nonintegrable real vector field to describe the scaling  under parallel transfer of quantities.  The problem, as noted by Einstein \cite{Weyl}, was that this required that the properties of measuring rods, clocks, and atomic spectra depend on their past history. He also noted that this contradicts the known physical properties of rods, clocks, and atomic spectra. A good description of Weyl's work, including Einstein's remarks and subsequent developments of early gauge theory are in a book by O'Raifeartaigh \cite{OR}.}  Scaling is an important component of renormalization theory \cite{Peskin}, scale invariance of physical quantities,  and of conformal field theories \cite{Ginsparg,Gaberdiel}.

            The number structure scaling used here  in the connections between fibers is different from these types of scaling. It differs from conformal transformations in that both angles between vectors and vector lengths are scaled.  The scaling of angles may seem strange and counterintuitive.   However, as will be seen, this does not cause problems. For example, the properties of number scaling are such that trigonometric relations are preserved under scaled parallel transfer from one fiber to another.

            The basic difference between the scaling used here and the other types of scaling used in physics is that both quantities and multiplication operations are scaled. The scaling must be such that the axiom validity of the  number and vector space structures is preserved under scaling. The type of scaling used here and in earlier work is closest to that in a recent paper in which functional relations between number structures of different types include  the basic operations as well as the quantities \cite{Czachor}.

            The plan of the paper is to first describe  the salient aspects of scaling for both number and vector space structures.  Because of its importance to the rest of this work, it is done  in some detail in Section  \ref{NS}.   Section \ref{FB} comes next with  a description of the parts of the theory of fiber bundles  that are needed here. The bundle fiber contains pairs of scalar structures and vector spaces at all levels of scaling.

            Connections between fibers are described in Section \ref{Cn}. A space or space time dependent scaling field  is introduced to describe the vertical component of connections. The effect of this connection field on both scalar and vector structure valued fields and on scalar and vector valued fields is described.

            The effect of number scaling fields on Lagrangians, actions, and equations of motion for Klein Gordon and Dirac fields is described next in Section \ref{GT}. Abelian gauge theory for pure number scaling  Lagrangians shows that the imaginary part of the gradient, $\vec{A}+i\vec{B}$, of the scaling field must have no mass.  The real part of the gradient can have any mass including $0.$ Inclusion of the electromagnetic field does not affect these mass conditions.

            The Lagrangian for the Mexican hat potential with the scaling field present is used in subsection \ref{HMABF} to find the effect of the Higgs mechanism on the scaling field.  The $\vec{B}$ field combines with a Goldstone boson to give a massive vector boson, a photon, and a Higgs field. The $\vec{A}$ field remains with any mass possible.  The very speculative possibility that $\vec{A}$ might be the gradient of the Higgs field is noted.

            The effect of scaling on some geometric quantities occupies Section \ref{NSG}. Both curve lengths and the geodesic equation  are derived.  It is seen that scaling affects the proper time of clocks carried along a path.

             Experimental restrictions on the possible physical properties of $\vec{A}$ and $\vec{B}$ are discussed in Section \ref{ERAB}. The lack of direct physical evidence for the presence of the $\vec{A}$ and $\vec{B}$ fields means that the values of these fields,  in a region of cosmological space and time occupiable by us as observers and by other intelligent beings with whom we can effectively communicate, must be too small to be observed. There are no restrictions on the values of these fields outside the region.   The region should be small with respect to the size of the observable universe.

            Section \ref{C} concludes the paper. It is noted that number scaling has no effect on comparisons of computations with one another or with results of measurements carried out at different space time points.

           It should be noted that the idea of local mathematical structures is not new.  It has been described in the context of category theory in which locality refers to structure  interpretations in different categories \cite{Bell}. Here locality refers to locations in space time. Locality in gauge field theories has also been discussed \cite{Mack}, but not with respect to mathematical structures.

            \section{Number scaling}\label{NS}
            A brief but explicit description of number scaling is in order as it is basic to all that follows.  One begins with the observation that mathematics is based on a collection of structures of different types of mathematical systems with relations or maps between them \cite{Gold,MM}.  A structure consists of a base set, and a few basic operations, relations, and constants for which a relevant set of axioms are true.  These are referred to as models in mathematical logic \cite{Barwise,Keisler}.

            Examples of structures include the different types of numbers, (natural integers, rational, real, complex),  vector spaces, algebras, etc.  The usual structures for the different types of numbers and the axioms sets they satisfy are listed below.
           \begin{itemize}
            \item $\bar{N}=\{N,+,\times,<, 0,1\}$
            Nonnegative elements of a discrete
            ordered commutative ring with identity
            \cite{Kaye}.

            \item $\bar{I}=\{I,+,-,\times,<, 0,1\}$
            Ordered integral domain \cite{integer}.

            \item $\overline{Ra}=\{Ra,+,-,\times,\div,<, 0,1\}$
            Smallest ordered field \cite{Rational}.

            \item $\bar{R}=\{R,+,-,\times,\div,<, 0,1\}$
            Complete ordered field \cite{real}.

            \item $\bar{C}=\{C,+,-,\times,\div,^{*}, 0,1\}$
            Algebraically closed field of characteristic $0$ plus
            axioms for complex conjugation \cite{complex}.
            \end{itemize}

            Here and in the following, an overline, such as in $\bar{N},$ denotes a structure. No overline, as for $N$, denotes a base set. The complex conjugation operation has been added as a basic operation to $\bar{C}$ as it makes the development much easier.

            As usually used, these structures conflate two distinct concepts, that of numbers as elements of the base sets, and that of number values acquired by the base set elements as part of a structure containing the base set. Natural numbers are the easiest set in which to see the distinction between number and number value.

            To understand this consider the even natural numbers.  These are just as good a structure for the natural numbers as are all the natural numbers.   The relevant structure is given by
            \begin{equation}\label{BN2}\bar{N}^{2}=\{N_{2},+_{2},\times_{2},>_{2},0_{2},1_{2}\}.\end{equation}Here $N_{2}$ is a subset of $N$ consisting of every other element of $N$.  An explicit example of $N_{2}$ is obtained from Von Neumann's characterization of the natural numbers in set theory as the sets, $\phi, \{\phi\}, \{\phi,\{\phi\}\},\cdots$ for $N.$ Here $N_{2}$ consists of the sets, $\phi,\{\phi,\{\phi\}\},\cdots.$ Here $\phi$ denotes the empty set.

            Comparison of the structures, $\bar{N}$ and $\bar{N}^{2}$  shows that the element of $N_{2}$ that has value $1_{2}$ in $\bar{N}^{2}$ has value $2$ in $\bar{N}.$ In general, the element of $N_{2}$ that has value $n_{2}$ in $\bar{N}^{2}$ has value $2n$ in $\bar{N}.$ The subscript $2$ on values denotes the structure to which the values belong.

            This is the simplest example of the distinction between numbers as base set elements and number values of the base set elements in structures. It extends to structures that are described by $\bar{N}^{n}$ where\begin{equation}\label{BNn}\bar{N}^{n}=\{N_{n},+_{n},\times_{n},<_{n}, 0_{n},1_{n}\}.\end{equation}This structure is based on the fact that the multiples of $n$ are also a model of the natural number axioms. Here one sees that an element $a$ of $N_{n}$ has a value in $\bar{N}^{n}$ that is $1/n$ times the value it has in $\bar{N}.$ If $2$ is a factor of $N$ then elements of $N_{n}$ are also elements of $N_{2}$, and an element $a$ of $N_{n}$ has a value in $\bar{N}^{n}$ that is $2/n$ times the value $a$ has in $\bar{N}^{2}.$ Conversely the value of an element $a$ (in $N_{2}$ and $N_{n}$) in $\bar{N}^{2}$ is $n/2$ times the value of $a$ in $\bar{N}^{n}.$

            This concept of value change or scaling can be extended to the other components of $\bar{N}^{n}$ to define a structure that expresses the values that $+_{n},\times_{n},$ and $<_{n}$ have in $\bar{N}^{2}$.  The resulting structure is defined by $\bar{N}^{n}_{2}$ where \begin{equation} \label{BN2n}\bar{N}^{n}_{2}=\{N_{n},+_{2},\frac{2}{n}\times_{2},<_{2},0_{2},\frac{n}{2}1_{2}\}.
            \end{equation}The scaling of multiplication  is not arbitrary.  It is done to satisfy the requirement that $\bar{N}^{n}_{2}$ satisfy the axioms of arithmetic if and only if $\bar{N}^{n},\bar{N}^{2},$ and $\bar{N}$ do.  Addition is unchanged. Note that addition and multiplication are operations on number values, not on base set numbers.

            The structure $\bar{N}^{n}_{2}$ describes a relative scaling between  the values of the elements of $N_{n}$ in $\bar{N}^{n}$ and $\bar{N}^{2}.$ This relative scaling can be extended to any pair of numbers $n,m$ where $m$ is a factor of $n.$  One substitutes $m$ for $2$ everywhere in the above description.

            The description of number scaling for the natural numbers applies with minor changes to number scaling for the other types of numbers. Scaled structures for the rational, real, and complex numbers are given by,\footnote{Integers are not included as the description is  essentially the same as that for the natural numbers.}\begin{equation} \label{BRas}\overline{Ra}^{s} =\{Ra, \pm_{s}, \times_{s}, (-)_{s}^{-1_{s}}, <_{s},0_{s},1_{s}\},\end{equation} \begin{equation}\label{BRs} \bar{R}^{s}=\{R,\pm_{s},\times_{s}, (-)_{s}^{-1_{s}},<_{s},0_{s},1_{s}\},\end{equation} \begin{equation}  \label{BCs}\bar{C}^{s}=\{C,\pm_{s},\times_{s}, (-)_{s}^{-1_{s}},(-)_{s}^{*_{s}},0_{s},1_{s}\}.\end{equation}Here $s$ is a rational, real, and complex scaling factor for each of the corresponding number types.

            These structures differ from those for the natural numbers in that the base set is the same for all scaling factors.  This is a consequence of the fact that these number types are all fields; they are closed under division.

              It is often useful to define a generic number structure,
             \begin{equation}\label{genS}\bar{S}^{s}=\{S,Op_{s},Rel_{s},K_{s}\}.\end{equation} Here $Op_{s},$ $Rel_{s},$ and $K_{s}$ are the sets of basic operations, relations, and constants for the structure of type $S.$  For the number types considered here, $S=N,Ra,R,C$ with $s$ correspondingly natural, rational, real or complex.

             Value maps of base set elements to values they have in structures are useful.  The value of an element, $a,$ in the base set $S$ in $\bar{S}^{s}$ is related to the value of $a$ in $\bar{S}^{t}$ by \begin{equation}\label{valta}val_{s}(a)=val_{s,t}(val_{t}(a))=\frac{t}{s} (val_{t}(a))_{s}. \end{equation}Here $(val_{t}(a))_{s}$ is the same number value in $\bar{S}^{s}$ as $val_{t}(a)$ is in $\bar{S}^{t}.$ The equation shows that if $a$ is such that $val_{t}(a)=1_{t},$ then $val_{s}(a)=(t/s)1_{s}.$

             It is important to keep the nomenclature clear.  Here and in the following, $a_{s}$ denotes a number value in $\bar{S}^{s}$.  Also $a_{s}$ is the same number value in $\bar{S}^{s}$ as $a_{t}$ is in $\bar{S}^{t}.$  However $a_{s}$ is \emph{not} the value the number, $a,$ as an element of the base set $S,$ has in $\bar{S}^{s}.$  This value is given by $val_{s}(a).$

            Let $t$ and $s$ be pairs of rational, real, and complex scaling factors.
            For rational, real, and complex  numbers, the structures corresponding to  $\bar{N}^{n}_{2}$ are
            \begin{equation}\label{BRaRCts}\begin{array}{c}\overline{Ra}^{t}_{s}=\{Ra,\pm_{s},\frac{\mbox{$s$}}
            {\mbox{$t$}}\times_{s}, \frac{\mbox{$t$}}{\mbox{$s$}}(-)^{-1_{s}},<_{s},0_{s},\frac{\mbox{$t$}} {\mbox{$s$}}1_{s}\}\\\\\bar{R}^{t}_{s}=\{R,\pm_{s},\frac{\mbox{$s$}}
            {\mbox{$t$}}\times_{s}, \frac{\mbox{$t$}}{\mbox{$s$}}(-)^{-1_{s}},<_{s},0_{s},\frac{\mbox{$t$}} {\mbox{$s$}}1_{s}\}\\\\\bar{C}^{t}_{s}=\{C,\pm_{s}, \frac{\mbox{$s$}}{\mbox{$t$}}\times_{s}, \frac{\mbox{$t$}}{\mbox{$s$}}(-)^{-1_{s}},\frac{\mbox{$t$}} {\mbox{$s$}} (-)^{*_{s}},0_{s},\frac{\mbox{$t$}}{\mbox{$s$}}1_{s}\}\end{array}\end{equation} The structure representations for rational and real number structures hold only if $t$ and $s$ are both positive or both negative.  If not, then $<_{s}$ is replaced by $>_{s}$ in the structures for $\overline{Ra}^{t}_{s}$ and $\bar{R}^{t}_{s}.$

             The rational, real, and complex number structures in Eq. \ref{BRaRCts} can be expressed generically by $\bar{S}^{t}_{s}$ where\begin{equation}\label{BSts}\bar{S}^{t}_{s}=\{S,\pm_{s}, \frac{s}{t}\times_{s},\frac{t}{s}(-)^{-1_{s}}, <_{s},0_{s}\frac{t}{s}1_{s}\}\end{equation}for $S=Ra$ and $S=R$. For $S=C$, $<_{s}$ is replaced by $(t/s)(-)^{*_{s}}.$

             The relation between structures, $\bar{S}^{t}$ with a superscript only and those such as $\bar{S}^{t}_{s}$ with both superscript and subscript show that $\bar{S}^{t}=\bar{S}^{t}_{t}.$  The subscript index labels the basic operations and relations in the structure. The relations between $\bar{S}^{t}$ and $\bar{S}^{t}_{s}$ can be made explicit by a map, $Z_{s,t}$, of the components of $\bar{S}^{t}$ onto the components of $\bar{S}^{t}_{s}$.  One has\begin{equation}\label{ZstS}Z_{s,t} \bar{S}^{t}=\bar{S}^{t}_{s}\end{equation} where\begin{equation}\label{Zst}\begin{array}{c} Z_{s,t}(a)=a \mbox{ for each $a$ in $S$,}\\\\Z_{s,t}(\pm_{t})=\pm_{s},\;\;\;Z_{s,t}(\times_{t})= \frac{\mbox{$s$}} {\mbox{$t$}}\times_{s},\;\;\; Z_{s,t}((-)^{-1_{t}})=\frac{\mbox{$t$}}{\mbox{$s$}}(-)^{-1_{s}}\;(S=Ra,R,C),\\\\ Z_{s,t}((-)^{*_{t}})=\frac{\mbox{$t$}}{\mbox{$s$}} (-)^{*_{s}}\;(S=C),\;\;\;Z_{s,t}(<_{t})=<_{s} \;(S=N,Ra,R),\\\\Z_{s,t}(0_{t})=0_{s},\;\;\;Z_{s,t}(1_{t}) =\frac{\mbox{$t$}}{\mbox{$s$}}1_{s}, \;\;\;Z_{s,t}(a_{t})=\frac{\mbox{$t$}}{\mbox{$s$}}a_{s}\mbox{ for all values, $a_{t}$}.\end{array}\end{equation}

             The properties of $Z_{s,t}$ show that it is the identity on the base set of numbers.  but number values are scaled by $t/s$,  The scaling of the operations is based on the following implications:$$a_{t}\pm_{t}b_{t}\rightarrow (\frac{t}{s}a_{s})\pm_{s}(\frac{t}{s}b_{s})=\frac{t}{s} (a_{s}+_{s}b_{s}),$$ $$a_{t}\times_{t}b_{t}\rightarrow(\frac{t}{s}a_{s})(\frac{s}{t}\times_{s}) (\frac{t}{s}b_{s})=\frac{t}{s} (a_{s}\times_{s}b_{s}),$$
            $$(a_{t})^{-1_{t}}\rightarrow \frac{t}{s}(a_{s}^{-1_{s}}).$$  The inverse does not scale as $(a_{t})^{-1_{t}}\rightarrow (t/s)((t/s)a_{s})^{-1_{s}}.$  This can be seen by replacing the inverse by the division operation where $\div_{t}\rightarrow (t/s)\div_{s}.$  One has $$a_{t}^{-1_{t}}=1_{t}\div_{t}a_{t}\rightarrow (\frac{t}{s}1_{s})(\frac{t}{s}\div_{s})(\frac{t}{s}a_{s})
           =\frac{t}{s}(1_{s}\div_{s}a_{s})=\frac{t}{s}a_{s}^{-1_{s}}.$$

            The scaling of the basic operations\footnote{Variation of operations in abstract algebras has been noted in passing in \cite{Bell}.} and number values indicated in Eq. \ref{Zst}  is not arbitrary.  It is done to ensure that the validity of the relevant axiom sets for the different number types is preserved under scaling.

            The fact that  axiom validity is invariant under scaling of number structures can be extended to properties in general.  If $P$ is a property of number values, then for each pair, $s,t$ of scaling factors,  and number value, $a_{s},$   $P^{s}(a_{s})$ is true for $\bar{S}^{s}$ if and only if $P^{t}(a_{t})$ is true for $\bar{S}^{t}$.  $P^{t}$ is obtained from $P^{s}$ by replacing all basic operations used to describe $P$ in $\bar{S}^{s}$ by the corresponding ones in $\bar{S}^{t}.$

            These aspects of number scaling emphasize the fact that the elements of the base set, $S,$ must be separated from the values they have in different structures. However, there is just one number where the distinction between elements of $S$ and their values in a structure is not needed. This is the number $0.$ This number has the same value, $0,$ in all scaled structures. It is like a number vacuum in that the value, $0$, is invariant under all scalings.

            The properties of number scaling are such that the $mth$ term  $T^{t}_{m}$ in a power series in $\bar{S}^{t}$ corresponds to $t/s)T^{s}_{m}$ in $\bar{S}^{t}_{s}$.  Here $T^{s}_{m}$ is the same number value in $\bar{S}^{s}$ as $T^{t}_{m}$ is in $\bar{S}^{t}.$ It follows from  this that if $f^{t}$ is an analytic function in $\bar{S}^{t}$ (for $S=R,C$) with values $f^{t}(a_{t})$,  then $(t/s)f^{s}$ is the corresponding analytic function in $\bar{S}^{t}_{s}$ with values $(t/s)f^{s}(a_{s})$ in $\bar{S}^{t}_{s}.$

            Another nice property  of number scaling is that equations are preserved. This follows from the equivalences of equations for general mathematical terms.\footnote{The proof of this property is done by inductive construction of terms from the basic atomic expressions,  such as $a_{t}\times_{t}b_{t}$ and $a_{t}+_{t}b_{t}.$}  Equations involving analytic functions are a good example.  If $f^{t}$ is an analytic function, then $$f^{t}(a_{t})=b_{t}\leftrightarrow \frac{t}{s}f^{s}((a_{t})_{s})=\frac{t}{s}(b_{t})_{s} \leftrightarrow f^{s}(a_{s})=b_{s}.$$ The lefthand equation is in $\bar{S}^{t}$, the middle one is in $\bar{S}^{t}_{s}$  and the  righthand one is  in  $\bar{S}^{s}.$ Also the number values, $a_{t},b_{t},$ are the same values in $\bar{S}^{t}$ as $(a_{t})_{s},(b_{t})_{s}$ are in $\bar{S}^{t}_{s}$  as $a_{s},b_{s}$ are in $\bar{S}^{s}.$

            Vector spaces are also affected by scaling of the fields on which they are based.  This is a consequence of the fact that vector spaces are always spaces over a scalar field.  As such they cannot be considered in isolation.  One begins with the usual representation of a normed vector space $\bar{V}$ \cite{Kadison} over the scalar field, $\bar{S}.$ \begin{equation}\label{BV}\bar{V}=\{V, \pm,\cdot,|-|,v\}.\end{equation} Here $\cdot$ denotes scalar vector  multiplication, and $|-|$ denotes the real valued norm of vectors. $v$ denotes an arbitrary vector.

            Vector spaces can also be associated with scaled number structures.  Let $\bar{S}^{t}$ be a scaled scalar field with $t$ the scaling factor. Here $S$ is usually $R$ or $C$ with $t$ a real or complex scaling factor.  $S=Ra$ is also possible with $t$ a rational factor.  The  components of the vector space structure associated with $\bar{S}^{t}$ are  given by \begin{equation}\label{BVt}\bar{V}^{t}=\{V,\pm_{t},\cdot_{t}, |-|_{t},v_{t}\}.\end{equation}

            The relation between the vectors $v$ and $v_{t}$ are defined by the requirement that $v$ has the same properties relative to $\bar{S}$ as $v_{t}$ has relative to $\bar{S}^{t}$. For example $|v_{t}|_{t}$ must be the same number value in $\bar{S}^{t}$ as $|v|$ is in $\bar{S}.$  If the vectors, $u_{t}$ and $v_{t}$ in $\bar{V}^{t}$ are related by $u_{t}=s_{t}\cdot_{t}v_{t}$, then the corresponding vectors in $\bar{V}$ are related by $u=s\cdot v,$ and conversely. Here $s$ is the same number value in $\bar{S}$ as $s_{t}$ is in $\bar{S}^{t}.$ If $u,v,w$ are three vectors in $\bar{V},$ then $w=u\pm v$ if and only if $w_{t}=u_{t}\pm_{t}v_{t}$ in $\bar{V}^{t}$.

            These conditions show that $u_{t},v_{t},$ and $w_{t}$ are the same vectors in $\bar{V}^{t}$ as $u,v,w$ are in $\bar{V}.$  Note that $\bar{V}$ and $\bar{S}$ are vector space and scalar structures where the scaling factor is $1.$

            One can also define the vector space associated with the scalar field $\bar{S}^{t}_{s}.$ It is defined by \begin{equation} \label{BVts}\bar{V}^{t}_{s}=\{V,\pm_{s},\frac{s}{t}\cdot_{s}, \frac{t}{s}|-|_{s},\frac{t}{s}v_{s}\}.\end{equation}Here $v_{s}$ is the same vector  in $\bar{V}^{s}$ as $v_{t}$ is in $\bar{V}^{t}$ as $(t/s)v_{s}$ is in $\bar{V}^{t}_{s}.$ Also $|v_{t}|_{t}$ is the same number in $\bar{S}^{t}$ as $|v_{s}|_{s}$ is in $\bar{S}^{s}$ as $(t/s)|v_{s}|_{s}$ is in $\bar{S}^{t}_{s}.$

            The scaling  maps  of the components of $\bar{C}^{t}$ and $\bar{V}^{t}$ onto those given by Eqs. \ref{BSts} and \ref{BVts} will be referred to here and in later sections as  correspondences. For example, $|v_{t}|_{t}$ corresponds to the number value $(t/s)|v_{s}|_{s}$   and $(a_{t})^{*_{t}}$ corresponds to $(t/s)a_{s}^{*_{s}}$ both in $\bar{S}^{t}_{s}$ given in  Eq. \ref{BSts}. The term, 'correspondence', will also be used to describe the relations between the components of $\bar{S}^{t}$ and $\bar{V}^{t}$ to those of $\bar{S}^{t}_{s},$ Eq. \ref{BSts}, and $\bar{V}^{t}_{s},$ Eq. \ref{BVts}.

            One sees from the above  that  operations do not commute with scaling.  In the above the norm operation was done before the correspondence as $$v_{t}\rightarrow\frac{t}{s}|v_{s}|_{s}.$$ Reversing the order of the operations gives $$v_{t}\rightarrow(t/s)v_{s}\rightarrow |(t/s)v_{s}|_{s}.$$   These results are not the same unless $t/s$ is a nonnegative real number. For each case the decision on which comes first, scaling or operation, is based on the requirement of axiom validity preservation.

            There is another definition of scaled vector space structures that satisfies the axiom validity preservation requirement. It is given by \begin{equation}\label{bVpts}
            \bar{V}'^{t}_{s}=\{V,\pm_{s},\frac{s}{t}\cdot_{s},\frac{t}{s}|-|_{s},v_{s}\}.\end{equation}  The difference between this structure and that of $\bar{V}^{t}_{s}$ in Eq. \ref{BVts} is that the vectors are not scaled.

            This structure is not used here because it fails the equivalence between $n$ dimensional vector spaces and their representations based on complex numbers.  Thus one has $\bar{V}\equiv \bar{C}^{n}$ and $\bar{V}^{t}_{s}\equiv (\bar{C}^{t}_{s})^{n},$ but not $\bar{V}'^{t}_{s}\equiv (\bar{C}^{t}_{s})^{n}.$

             \section{Fiber bundles}\label{FB}
               Fiber bundles have been much used to describe  gauge theories and other areas of physics and geometry \cite{Manin}-\cite{Pfeifer}. Here they will be used  to describe the effects of number scaling on gauge theories and on some geometric objects. A fiber bundle \cite{Husemoller,Husemoller2} can be described by $\mathcal{B}=\{E,\pi,M \}.$ Here $E$ and $M$ are the total and base spaces,  and $\pi$ is a projection of $E$  onto $M$. For each point, $x,$ in $M$, $\pi^{-1}(x)$ is a fiber, at $x$, in $E$.

               A fiber bundle is a product bundle if the total space  is a product as in $E=M\times F.$ Here $F$ is a fiber, and $\pi^{-1}(x)=(x,F)$ for all $x$ in $M$. Here the interest is in product bundles because $M$ is flat as Euclidean or Minkowski spaces.

               Here fibers are taken to be products of vector spaces and scalar structures on which the vector spaces are based. The corresponding product bundle is given by \begin{equation}\label{MFSV}
               \mathfrak{SV}=M\times (\bar{S}\times\bar{V}),\pi,M.\end{equation}This bundle is a fiber product \cite{Husemoller} of the bundles $M\times \bar{S},\pi,M$ and $M\times \bar{V},\pi,M.$

               The fiber at each point, $x,$ of $M$ in the bundle $\mathfrak{SV}$ is given by \begin{equation} \label{pim1x}\pi^{-1}(x)=x\times\bar{S}\times\bar{V}=\bar{S}_{x}\times\bar{V}_{x}.\end{equation}Both $\bar{S}_{x}$ and $\bar{V}_{x}$  are local mathematical structures in that they are associated with a point $x$ of $M$.  The bundle is a good description of local scalar and vector space structures associated with each point of $M$.

               The fiber bundle based representation of local mathematical structures will be used throughout this work. Gauge theories make use of this representation \cite{Daniel,Drechsler}. For these theories fibers at each point of $M$   contain gauge groups as structure groups acting on vector spaces. Principal frame bundles of vector space bases  are associated with the vector space fiber bundles.   However the scalar field associated with the vector spaces is taken to be a single complex number field not associated with any point of $M$.

               This is changed here by associating a scalar structure with the vector space at each point of $M$. This is shown in the description of $\mathfrak{SV}$ where the fiber at $x$ contains both local scalar and vector space structures, $\bar{V}_{x}$  and $\bar{S}_{x}.$ The components of $\bar{V}_{x}$ include $|-|_{x}$ as a map from $\bar{V}_{x}$ to $\bar{S}_{x}$ and $\cdot_{x}$ as  a map from $\bar{S}_{x}\times \bar{V}_{x}$ to $\bar{V}_{x}.$

                A main point of this work is to use fiber bundles to describe the effect of number scaling on gauge theories and on some geometric objects.  This can be done by expanding the fibers to include all scaled pairs of scalar fields and vector spaces. For gauge theories with complex scalars the pairs are $\bar{C}^{c}\times\bar{V}^{c}$ for all complex numbers. $\bar{C}^{c}$ and $\bar{V}^{c}$ are given by Eqs. \ref{BCs} and \ref{BVt}. For geometric objects the pairs are $\bar{R}^{r}\times \bar{T}^{r}.$ $\bar{R}^{r}$ is a scaled real number structure and $\bar{T}^{r}$ is a scaled tangent space.

                Inclusion of number scaling in a fiber bundle description of gauge theories can be achieved by first defining fiber bundles, \begin{equation}\label{mcCVc}\mathfrak{CV}^{c}=\{M\times (\bar{C}^{c}\times\bar{V}^{c}),p^{c},M\}
                \end{equation} for all complex scaling factors, $c$.  Here $\mathfrak{CV}^{c}$ is a fiber product as defined by Eq. \ref{MFSV}.

                One then defines a sum bundle by\begin{equation}\label{sumcCV}\mathfrak{CV}^{\cup}=\sum_{c}\mathfrak{CV}^{c}=\{M\times \bigcup_{c}(\bar{C}^{c}\times\bar{V}^{c}),Q,M\}.\end{equation} Here $\bigcup_{c}(\bar{C}^{c}\times\bar{V}^{c})$ is the collection of all  pairs, $(\bar{C}^{c}, \bar{V}^{c}).$ $Q$ projects the total space onto $M$.  The inverse map $Q^{-1}$ is defined by \begin{equation}\label{Qm1x}Q^{-1}(x)=(x,\bigcup_{c}(\bar{C}^{c}\times\bar{V}^{c})) =\bigcup_{c}(\bar{C}^{c}_{x}\times\bar{V}^{c}_{x})\end{equation} Here $\bar{C}^{c}_{x}$ and $\bar{V}^{c}_{x}$ are defined from $(x,\bar{C}^{c})$ and $(x,\bar{V}^{c})$ as was done  in Eq. \ref{pim1x}. The projection $Q$ is related to the individual projections, $p^{c}$ by \begin{equation}\label{QxbC}Q(x,(\bar{C}^{c}\times\bar{V}^{c})) =p^{c}(x,(\bar{C}^{c}\times\bar{V}^{c})). \end{equation}

                All complex number structures, $\bar{C}^{c}_{x},$ in the fiber at $x$ have the same base set, $C_{x}.$  All possible number values for each  number in $C_{x}$  are included in structures in $\mathfrak{CV}^{\cup}.$. For each number $a$ in $C_{x}$ and number value $q$, there is a scaled  complex number structure in $\mathfrak{CV}^{\cup}$ in which $a$ has $q$ as a value. The vector space structures for all $c$ at $x$  also have the same base set, $V_{x}$.\footnote{At this point it is not clear if base sets in the structures in the fibers at the different points of $M$ need to be distinguished by a subscript, $x$. What is clear is that they must be the same for all values of $c$ in a fiber.}  These results follow from Eqs. \ref{BCs}, \ref{BVt}, and \ref{pim1x}.

                A structure group, $W_{CV}$ with elements, $W_{CV,d},$ that are structure isomorphisms for any complex number $d,$  can be defined on the fiber. The action of $W_{CV,d}$ on $\bigcup_{c}(\bar{C}^{c}\times\bar{V}^{c})$  is given by\footnote{$0$ can be included as a value of $d$ if empty or $0$ structures are allowed. Then the action of $W_{CV,0}$ on a structure empties or annihilates it.} \begin{equation}\label{WCVd}W_{CV,d}\bigcup_{c}(\bar{C}^{c}\times\bar{V}^{c})=\bigcup_{c} (\bar{C}^{cd}\times\bar{V}^{cd}).\end{equation}The group $W_{CV}$ is a continuous commutative group with $W_{CV,d^{-1}}$ the inverse of $W_{CV,d}.$ The scaling factors, $d,$ are all elements of the group, $GL(1,C).$ $W_{CV}$ also is the structure group for the fibers at each point, $x,$ of $M$.

                The bundle, $\mathfrak{CV}^{\cup}$ is a principal fiber bundle. The reason is the group $W_{CV}$ acts freely and transitively on the fibers in $\mathfrak{CV}^{\cup}$ \cite{Daniel,Drechsler}. No scaled structure is preferred over another.  All are completely equivalent.

                The definition of fiber bundles is quite general in that the fibers can contain much material.  For example a fiber can include a scalar field structure, $\bar{S},$ and   the mathematical analysis based on $\bar{S}.$  A fiber can include integrals and derivatives of functions, charts of $M$ and other systems. In all cases the fiber bundle is represented here by $\{M\times F,\pi,M\}.$  The bundle is trivial on $M$ because $M$ is flat as Minkowski or Euclidean space.   The range set of $\pi^{-1}$ consists of local structures at all points of $M$. This generalization will be made use of in the following sections.

                \section{Connections}\label{Cn}

                Connections or parallel transports \cite{Mack} play important roles in gauge theories and geometry.  They account for the fact that, for vector spaces at different points of $M$,  a choice of a basis  at point $y$ does not determine the choice  at point, $x$ of $M$.  This is the "Naheinformationsprinzip", 'no information at a distance' principle \cite{Mack}.  Applied to number scaling it says that the choice of a number scaling factor at point $y$ does not determine the choice at point $x.$

                Connections  enable one to compare values of quantities in  fibers at different $M$ locations.  These are needed to describe  physical quantities such as derivatives and integrals on $M$.    They connect a quantity at an arbitrary scaling  level  in a fiber at $x$ to a quantity at a neighboring level in a fiber at a neighboring point of $M$. The connection can be decomposed into two components, a horizontal one that connects a quantity at point $x$ to a quantity at the same level but at point $x+\vec{dx}$ and a vertical component that connects a quantity at point $x$ to a quantity at a different level at point $x$.

                 Connections can be used to map  scalar and vector structures in a $\mathfrak{CV}^{\cup}$ bundle fiber at $x+\vec{dx}$ to a structure in a fiber at $x.$ These are the simplest to understand. and are discussed first. The results are used to  map number values and vectors inside  structures in a fiber at $x$ to number values and vectors in structures in a fiber at neighboring points of $M$.

                 The number scaling connection is provided by a  smooth  $GL(1,C)$ valued field, $g$, on $M$. For each $x,$ $g(x)$ is a complex scaling factor. The values of $g$ are not elements of any complex scalar structure in a fiber.  In this sense they are global in that they apply equally to all fibers. The field $g$ also acts at all levels, $c,$ of structures in the fibers in $\mathfrak{CV}^{\cup}.$

                 \subsection{Scalar and vector structure valued fields}\label{SVS}

                 To see how the connection works it is useful to first consider  a structure valued field $\Psi_{cg}.$ Here $c$ is an arbitrary level in the fiber bundle. For a complex number structure valued field, \begin{equation}\label{Psicx} \Psi_{cg}(x)=\bar{C}^{cg(x)}_{x}\end{equation}for each $x$ in $M$.  For a vector structure valued field,\begin{equation}\label{PsicVx} \Psi_{cg}(x)=\bar{V}^{cg(x)}_{x}.\end{equation} These definitions show that for each $c$ and $g,$ $\Psi_{cg}$ is unique in that for each $x$ just one scalar or vector structure value is possible for $\Psi_{cg}(x).$  This simplifies the description of  field quantities such as derivatives.

                 The derivative of $\Psi_{cg}$ makes use of the connections. The components of the derivative of $\Psi_{cg}(x)$ have the form \begin{equation} \label{DmuxP}D_{\mu,x}\Psi_{cg}=\frac{\Psi_{cg}(x+d^{\mu}x)- \Psi_{cg}(x)}{d^{\mu}x}.\end{equation}The limit $d^{\mu}x\rightarrow 0$ is implied.  Replacing $\Psi_{cg}(x)$ and $\Psi_{cg}(x+d^{\mu}x)$ by their structure values gives \begin{equation} \label{DmuxcS}D_{\mu,x}\Psi_{c}=\frac{\bar{C}^{cg(x+d^{\mu}x)}_{x+d^{\mu}x}- \bar{C}^{cg(x)}_{x}}{d^{\mu}x}.\end{equation} The implied subtraction in the numerator does not make sense because structure subtraction can be defined  only between structures at the same level within a fiber.  It is not defined between structures at different levels in a fiber or between structures at the same level in different fibers.

                 Connections are used here to map $\bar{C}^{cg(x+d^{\mu}x)}_{x+d^{\mu}x}$ to a structure in the fiber at $x$ and at the same level as is $\bar{C}^{cg(x)}_{x}.$  The  horizontal component of the connection maps $\bar{C}^{cg(x+d^{\mu}x)}_{x+d^{\mu}x}$ to the same structure, $\bar{C}^{cg(x+d^{\mu}x)}_{x},$ in the fiber at $x$ and the vertical component changes the level of $\bar{C}^{cg(x+d^{\mu}x)}_{x}$ to $cg(x)$. The subtraction can then be carried out as both structures are at the same level in the fiber at $x.$

                 The horizontal component of the connection is taken here to be the identity.\footnote{In other work \cite{SPIE5}, the horizontal component  included a complex vector field, whose  integrability was not known.}  As a result, the derivative  in Eq. \ref{DmuxcS} can be written as\begin{equation}\label{DmuxcC}D_{\mu,x}\Psi_{cg}= \frac{\bar{C}^{cg(x+d^{\mu}x)} _{x}- \bar{C}^{cg(x)}_{x}}{d^{\mu}x}.\end{equation}The vertical connection component makes use of the number scaling map in Eq. \ref{BSts} where $S=C.$ One also makes use of the fact that for any fiber level, $t$, $\bar{C}^{t}=\bar{C}^{t}_{t}.$  The result is given by \begin{equation}\label{DmuxcCc}D_{\mu,x}\Psi_{cg}= \frac{\bar{C}^{cg(x+d^{\mu}x)} _{cg(x),x}- \bar{C}^{cg(x)}_{cg(x),x}}{d^{\mu}x}= \frac{\frac{d} {e}\bar{C}^{e}_{e,x}-\bar{C}^{e}_{e,x}} {d^{\mu}x}.\end{equation} Here\begin{equation} \label{BCcgcg} \bar{C}^{d} _{e,x}=\frac{d}{e}\bar{C}^{e}_{e,x}\end{equation}has been used. To save on notation $cg(x+ d^{\mu}x)$ and $cg(x)$ have been replaced respectively by $d$ and $e$.

                 The components of $\frac{d}{e} \bar{C}^{e}_{e,x}$ are obtained from Eq. \ref{BSts}. They are given by\begin{equation}\label{BCdex} \bar{C}^{d}_{e,x} =\{C_{x},\pm_{e},\frac{e}{d} \times_{e},\frac{d}{e}(-)^{-1_{e}}, \frac{d}{e}(-)^{*_{e}}, 0_{e},\frac{d}{e}1_{e}\}=\frac{d}{e}\bar{C}^{e}_{e}.\end{equation} The subscript, $x,$ is left off of the operations and number values in $\bar{C}^{d}_{e}.$

                 Eq. \ref{BCcgcg} enables one to combine the terms in the numerator of the derivative.  The result is \begin{equation}\label{DPsicg}D_{\mu,x}\Psi_{cg}=\frac{(\frac{d}{e}-1) \bar{C}^{e}_{e,x}}{d^{\mu}x}.\end{equation}Replacing $d$ and $e$ by their equivalents, $cg(x+d^{\mu}x)$ and $cg(x)$, and using a Taylor expansion of $g(x+d^{\mu}x)$ gives
                 \begin{equation}\label{DxP}D_{\mu,x}\Psi_{cg}=\frac{c\partial_{\mu,x}(g)d^{\mu}x}{cg(x)d^{\mu}x} \bar{C}^{cg(x)}_{cg(x),x} =\frac{\partial_{\mu,x}(g)}{g(x)}\Psi_{cg}.\end{equation} This shows that the effect of number scaling on structure field derivatives is independent of the scaling level, $c,$ in the fiber bundle. The level cancels out because the ratio, $d/e,$ or its inverse, appear as scaling factors for the structure components. $d$ and $e$ do not appear separately.

                 Eq, \ref{DxP} also holds if $\Psi_{cg}$ is a vector space structure valued field.  This follows from the fact that the same scaling factor ratio shows in the components of the scaled structure $\bar{V}^{d}_{e}$, Eq. \ref{BVts}, as in $\bar{C}^{d}_{e}.$

                 \subsection{Scalar and vector valued fields}\label{SV}
                 The results obtained for structure valued fields can be taken over to fields as sections on the fiber bundle, $\mathfrak{CV}^{\cup}.$  Let $\psi_{cg}$ be a vector valued field where for each $x$ in $M$, $\psi_{cg}(x)$ is a vector in $\bar{V}^{cg(x)}_{x}$ in the fiber at $x.$ One proceeds  in a similar fashion to that used for the structure valued fields.  The partial derivative is given by \begin{equation}\label{pdmux}\partial_{\mu,x}\psi_{cg}=\frac{\psi_{cg} (x+d^{\mu}x)-\psi_{cg}(x)}{d^{\mu}x}.\end{equation} The first and second number terms are vectors in $\bar{V}^{cg(x+d^{\mu}x)}_{x+d^{\mu}x}$ and $\bar{V}^{cg(x)}_{x}.$

                 One first uses the horizontal component of the connection to map $\psi_{cg}(x+d^{\mu}x)$ into the same  vector, $\psi_{cg}(x+d^{\mu}x)_{x}$ in $\bar{V}^{cg(x+d^{\mu}x)}_{x}$ as $\psi_{cg}(x+d^{\mu}x)$ is in $\bar{V}^{cg(x+d^{\mu}x)}_{x+d^{\mu}x}.$  The vertical component  is then used to map the result into the same vector in $\bar{V}^{cg(x+d^{\mu}x)}_{cg(x),x}.$ The map sequence gives the vectors, \begin{equation} \label{psicgx}\psi_{cg}(x+d^{\mu}x)\rightarrow \psi_{cg}(x+d^{\mu}x)_{x} \rightarrow\frac{cg(x+d^{\mu}x)}{cg(x)}\psi_{cg}(x+d^{\mu}x)_{x}.\end{equation}

                 Replacement of $\psi_{cg}(x+d^{\mu}x)$ in Eq. \ref{pdmux} by the right hand term of Eq. \ref{psicgx} and use of a  Taylor expansion of $g(x+d^{\mu}x)$  gives
                 \begin{equation}\label{Dmup}D_{\mu,x}\psi_{cg}=(\partial_{\mu,x}+\frac{\partial_{\mu,x}g} {g(x)})\psi_{cg}(x)\end{equation} as the result.  This differs from the derivative for the structure valued field in that a partial derivative is also present and nonzero in general. Also, as was the case for $\Psi_{cg}$, the result is independent of $c.$

                This expression can be put in a more recognizable form by expressing the scaling field $g$ as the exponential of a pair of scalar fields as in \begin{equation}\label{gxe}g(x)= e^{\alpha(x)+i\beta(x)}.\end{equation} This gives the final result,\begin{equation} \label{Dmuptp}D_{\mu,x}\psi_{cg}=(\partial_{\mu,x}+ A_{\mu}(x)+iB_{\mu}(x)) \psi_{cg}. \end{equation} Here $\vec{A}$ and $\vec{B}$ are the gradients of $\alpha$ and $\beta.$  This replacement of $\partial_{\mu,x}g/g$ by $A_{\mu}(x)+iB_{\mu}x$ also applies to the derivatives of the structure valued field, $\Psi_{cg}.$

            \section{Gauge theories}\label{GT}
            The presence of number scaling extends the reach of gauge theories by adding another gauge group, $GL(1,C)$, to those already used in the standard model and other areas of physics. The simplest case to consider is that in which $GL(1,C)$ is the only gauge group.  This is an Abelian theory for pure number scaling.

            The covariant derivative is given by Eq. \ref{Dmuptp}.  However it is instructive to obtain the derivative in another way well known \cite{Montvay} in gauge theory.  The covariant derivative is given by\begin{equation}\label{DmuxY}D_{\mu,x}\psi=\frac{Y(x,x+d^{\mu}x)\psi(x+d^{\mu}x)- \psi(x)}{d^{\mu}x}.\end{equation}Here $Y(x,x+d^{\mu}x)$ is an element of the group $GL(1,C)$. It accounts for the freedom of choice of number scaling factors in the mapping of $\psi_{cg}(x+d^{\mu}x)$ in the fiber at $x+d^{\mu}x$ to the fiber at $x.$ Setting \begin{equation}\label{Yxxdx}Y(x,x+d^{\mu}x)= e^{(A_{\mu}(x)+iB_{\mu}(x))d^{\mu}x}\end{equation} and expanding the exponential to first order in small quantities gives Eq. \ref{Dmuptp}.

            This approach is quite different from that used above to obtain Eq. \ref{Dmuptp}. Eq. \ref{Yxxdx} introduces the vector fields, $\vec{A} +i\vec{B}$ directly.  There is no connection, implied or otherwise, of these fields to the gradient of a scalar field such as $g.$ As a result, in this approach, it an open question whether none, one, or both of these fields are integrable or not.  If the $\vec{B}$ field is nonintegrable, then the only way to distinguish it from the electromagnetic field is to give it a different coupling constant in Lagrangians.

            For these and other reasons the field, $\vec{A} +i\vec{B}$, is assumed here to be the gradient of a complex scalar field, $g.$  This greatly simplifies expressions and derivations.

             Here and from now on the subscripts $cg$ on the field, $\psi,$ are suppressed.  The reason is that the quantities such as covariant derivatives are independent of the value of $c.$  The descriptions can be shifted, with no change, from one level to another in $\mathfrak{CV}^{\cup}.$ All levels are equivalent.

            \subsection{Klein Gordon fields}\label{KGF}

            Number scaling affects the Klein Gordon Lagrangian \cite{Peskin} in that the partial derivatives are replaced by $D_{\mu,x}$ of Eq. \ref{Dmuptp}. In this case the Lagrangian density is given by\begin{equation} \label{LKGx}\mathcal{L}_{KG}(x)=|D_{\mu,x} \psi|^{2}-m^{2}\psi^{*}\psi.\end{equation}Expansion of the covariant derivative by use of Eq. \ref{Dmuptp} gives \begin{equation}\label{MLX} \begin{array}{l} \mathcal{L}_{KG}(x)=\partial_{\mu}\psi\partial^{\mu} \psi^{*}+a_{a}A_{\mu}(x)\partial^{\mu} (\psi\psi^{*})+ia_{b}B_{\mu}(x)(\psi\partial^{\mu}\psi^{*}-\psi^{*} \partial^{\mu}\psi)\\\\\hspace{2cm} +(a_{a}^{2}A_{\mu}(x)A^{\mu}(x)+a_{b}^{2}B_{\mu}(x)B^{\mu}(x)-m^{2})\psi^{*}\psi.\end{array} \end{equation} Coupling constants, $a_{a}$ and $a_{b},$ have been added. The second and third terms describe interactions of the $\vec{A}$ and $\vec{B}$ fields with the matter field, $\psi.$ The presence of $a_{a}^{2}A_{\mu}(x)A^{\mu}(x)+a_{b}^{2}B_{\mu}(x)B^{\mu}(x)$ in the mass term shifts the square of the mass  from $m^{2}$ to $m^{2}-a_{a}^{2}A_{\mu}(x)A^{\mu}(x)-a_{b}^{2}B_{\mu}(x)B^{\mu}(x).$

            This shows $\vec{A}$ and $\vec{B}$ as external fields that interact with the bosonic field, $\psi.$  The fact that these fields can  depend on the location, $x$ in $M,$ has consequences.  For example the shift of mass squared can depend on $x.$ This means that the mass of the field, $\psi,$ can depend on location.

            Another point to note is that the Lagrangian density. $L_{KG}(x),$ is a quantity in the fiber of $\mathfrak{CV}$ at $x.$ In this sense it is local. The density $L_{KG}(y)$ at another point $y$ of $M$ is in the fiber at $y$.

            The presence of the scalar fields, $\alpha(x)$ and $\beta(x)$ and their gradients affect the derivation of the equation of motion by minimizing the action.   The action  is an integral over all points of $M$ of $L_{KG}(x)$ as  in \begin{equation}\label{Ac}S=\int L_{KG}(x)d^{4}x.\end{equation} The integral is not defined as it corresponds to addition of integrands in different fibers of the bundle.

            This can be remedied by parallel transforming the integrands to values in a fiber at a reference point and then  defining the integration within a fiber.  This requires that the bundle $\mathfrak{CV}^{\cup}$  defined in Eq. \ref{sumcCV} be expanded to include a representation of $M$. The resulting bundle is given by \begin{equation}\label{sumcCVga} \mathfrak{MCV}^{\cup}= \sum_{c}\mathfrak{CV}^{c}=\{M\times\bigcup_{c} (\gamma(M)^{c}\times\bar{C}^{c}\times\bar{V}^{c}) ,Q,M\}.\end{equation}  Here $\gamma(M)^{c}$ is a chart representation of $M$ onto $(\mathbb{R}^{c})^{4}$, or $(\mathbb{R}^{c})^{3}$ for Euclidean space. The chart map domain can be all of $M$ because  is sufficient because $M$ is flat.

            The contents of a fiber at $x$ are now $\bigcup_{c}(\bar{C}^{c}_{x}\times\bar{V}^{c}_{x}\times \gamma(M)^{c}_{x})$ with $n=4$ or $n=3.$ Here $\gamma(M)^{c}_{x}=(\mathbb{R}^{c}_{x})^{n}.$ This expansion enables one to define space or space time integrals locally  as integrals over $(\mathbb{R}^{c}_{x})^{n}$ within a fiber at $x.$

            The reason one works with a representation of $M$ at level $c$ is that all factors in the integrand have to be at the same level in the fiber. The multiplication of $d^{4}x$ and $L_{KG}(x)$ in Eq. \ref{Ac} is defined between factors at the same fiber level. It is not defined for factors at different levels.

            Parallel transforming an integrand from $y$ to a reference point,  $x,$ must account for the fact that the integrand (including $d^{4}x$) has  real values at levels $cg(y)$ and $cg(x)$ in fibers at $y$ and $x.$ The summation over infinitesimal quantities implied by the integration is defined only for quantities at the same level in the fiber at $x.$ This requires that the $y$ integrand, which is a value in $\bar{R}^{cg(y)}_{x},$ must be scaled to to a value at level $cg(x)$. This level change map, $\bar{R}^{cg(y)}_{x}\rightarrow\bar{R}^{cg(y)}_{cg(x),x},$ corresponds to multiplying the integrand by the scaling factor\footnote{The parallel transform is independent of the path between $y$ and $x$ because $\lambda$ is a complex scalar field.} $$cg(y)/cg(x)=g(y)/g(x)=e^{\lambda(y)-\lambda(x)}.$$ Here Eq. \ref{gxe} is used along with $\lambda(x)=\alpha(x)+i\beta(x).$

            The resulting level $cg(x)$ expression for the action is \begin{equation}\label{Ax}S_{x}=e^{-\lambda(x)}\int_{x}e^{\lambda(y)}L_{KG}(y)_{x}d^{4}y_{x}. \end{equation}The integral is defined over $(\mathbb{R}^{cg(x)}_{x})^{4}.$

            The equation of motion is obtained in the usual way be setting the variation of the action equal to $0$.  The $\lambda$ field is considered to be a fixed external field for the variation. The Euler Lagrange equations and Eq. \ref{MLX} give \begin{equation}\label{KGEq}\begin{array}{l}\partial_{\mu,x}\partial^{\mu}_{x}\psi + (a_{a}\partial^{\mu}_{x}A_{\mu}(x)+ia_{b}\partial^{\mu}_{x}B_{\mu}(x))\psi+2ia_{b}B_{\mu}(x) \partial^{\mu}_{x}\psi\\\\+(m^{2}-a_{a}^{2}A_{\mu}(x)A^{\mu}(x)-a_{b}^{2}B_{\mu}(x)B^{\mu}(x))\psi
            \\\\+(\partial^{\mu}_{x}\lambda)(\partial_{\mu,x}\psi+(a_{a}A_{\mu}(x)+ia_{b}B_{\mu}(x))\psi)=0.\end{array}
            \end{equation}Here $(\partial^{\mu}_{x}\lambda)=A^{\mu}(x)+iB^{\mu}(x).$  The last line of this equation is the contribution of the $\exp(\lambda(y))$ factor to the equation of motion.

            This equation is quite complicated.  Finding a solution is difficult if not impossible.  The usual equation is obtained if either the $\vec{A}$ and $\vec{B}$ fields or the coupling constants are sufficiently small to be neglected. However the latter case still leaves the term $\partial^{\mu}\lambda\partial_{\mu}\psi$ in the equation.

            \subsection{Dirac Fields}\label{DF}

            The effect of number scaling  is also present for Dirac fields. This is accounted for by replacing $\partial_{\mu}$ in the  Dirac Lagrangian density \cite{Peskin}  by $D_{\mu}.$ The result is given by \begin{equation}\label{LDy} L_{D}(y)=\bar{\psi}i\gamma^{\mu}D_{\mu,y} \psi-m\bar{\psi}\psi= \bar{\psi}(i\gamma^{\mu}(\partial_{\mu,y}+a_{a}A_{\mu}(y)+ia_{b}B_{\mu}(y))-m)\psi.\end{equation} Coupling constants $a_{a}$ and $a_{b}$ have been added. Here $\bar{\psi}=\gamma^{0}\psi^{\dag}.$ In the bundle, $\mathfrak{CV}^{\cup},$ the Lagrange density is the numerical value of a quantity at level $cg(y)$ in the fiber at $y.$

            The Dirac equation is obtained by varying the action, \begin{equation}\label{SD}
            S_{D}=e^{-\lambda(x)}\int_{x}e^{\lambda(y)}L_{D}(y)d^{4}y.\end{equation} Here the integrands for the different points, $y$ are parallel transformed with scaling to a common reference point $x$ of $M$. The effects of scaling are accounted for by the exponential factors.

            Variation with respect to $\bar{\psi}$ gives, via the Euler Lagrange equations, \begin{equation} \label{DEq} i\gamma^{\mu} D_{\mu,x}\psi-m\psi=(i\gamma^{\mu}(\partial_{\mu,x}+a_{a}A_{\mu}(x)+ia_{b} B_{\mu}(x))-m)\psi=0.\end{equation}Variation with respect to $\psi$ gives a different equation because the exponential factor enters into the variation.  The Euler Lagrange equations are $$e^{\lambda(y)}\bar{\psi}i\gamma^{\mu} (a_{a}A_{\mu}+ia_{b}B_{\mu})-m) -\partial_{\mu,y}(e^{\lambda(y)}\bar{\psi}i\gamma^{\mu})=0.$$  Carrying out the indicated derivative gives \begin{equation}\label{DEB}(\partial_{\mu}+(1-a_{a})A_{\mu}+(1-a_{b})iB_{\mu}+m\bar{\psi} i\gamma^{\mu} =0.\end{equation}The replacement of $a_{a}$ and $a_{b}$ by $1-a_{a}$ and $1-a_{b}$ is due to the presence of the $e^{\gamma}$ factor in the action.

            The relationship of this equation to that of Eq. \ref{DEq} is of interest.  If $a_{a}=0=a_{b},$  then Eqs. \ref{DEq} and \ref{DEB} become respectively, \begin{equation}\label{aaab0}
            \begin{array}{c}i\gamma^{\mu}(\partial_{\mu,x}-m)\psi=0\\\\(\partial_{\mu}+A_{\mu}+iB_{\mu}+m)
            \bar{\psi} i\gamma^{\mu} =0.\end{array}\end{equation}If $a_{a}=1=a_{b}$ then the two equations become,\begin{equation}\label{aaab1}\begin{array}{c} i\gamma^{\mu} (\partial_{\mu,x}+A_{\mu}(x)+i B_{\mu}(x)-m)\psi=0\\\\(\partial_{\mu}+m)\bar{\psi} i\gamma^{\mu} =0.\end{array}\end{equation}

            \subsection{Abelian gauge theory}\label{Agt}

             In Abelian gauge theory  one requires that all the terms of the Lagrangians are invariant under the action of local $U(1)$ gauge transformations. This requirement applies to the Lagrangians in that the effects of number scaling are  expressed by the action of elements in the gauge group, $GL(1,C)$.  This group can be expressed as the product of $GL(1,R)$ and $U(1).$  The expression for the scaling function $g,$ as in Eq. \ref{gxe}, shows this explicitly.

            As is well known \cite{Cheng}, the restriction of  Lagrangians to local $U(1)$ invariant terms is done by the  replacement of  $D_{\mu,x}$ by $D^{\prime}_{\mu,x}$ where \begin{equation} \label{UDDU}D^{\prime}_{\mu,x}U(x)=U(x)D_{\mu,x}.\end{equation}Application of this condition to the Dirac Lagrangian of Eq. \ref{LDy} gives \begin{equation}\label{ABpAB}\begin{array} {c}A^{\prime}_{\mu}(x)=A_{\mu}(x)\\\\B^{\prime}_{\mu}(x) =B_{\mu}(x)+\frac{\mbox{$\partial_{\mu,x} a(x)$}}{\mbox{$a_{b}$}}.\end{array}\end{equation} Here $U(x)=e^{-ia(x)}$ and $a_{b}$ is the $B$-matter field coupling constant. It follows from this that the $\vec{B}$ field has zero mass.  Any mass including $0$ is possible for the $\vec{A}$ field.

            So far the effects of number scaling only in have included in the Lagrangians.
            Inclusion of the electromagnetic field requires expansion of the covariant derivative of Eq. \ref{DmuxY} to \begin{equation}\label{DmuxYU}D_{\mu,x}\psi=\frac{Y(x,x+d^{\mu}x) U(x,x+d^{\mu}x)\psi(x+d^{\mu}x)- \psi(x)}{d^{\mu}x}.\end{equation}Here $U(x,x+d^{\mu}x)=e^{iP_{\mu}(x)d^{\mu}x}$ where $\vec{P}$ is the photon field. Expansion of the exponentials to first order  in $d^{\mu}x$ gives \begin{equation}\label{DU1NS} D_{\mu,x}\psi=(\partial_{\mu,x}+ a_{a}A_{\mu}(x)+ia_{b}B_{\mu}(x) +ia_{p}P_{\mu}(x))\psi. \end{equation} for the covariant derivative. Here $a_{p}$ is the photon matter field coupling constant. Also $\psi$ is the electron field.

            The requirement that the terms in the Dirac Lagrangian be invariant under local $U(1)$ transformations  must account for the fact that the gauge group is \begin{equation}\label{GL1CU}
            GL(1,C)\times U(1)=GL(1,R)\times U(1)_{B}\times U(1)_{P}.\end{equation}  The subscripts $B$ and $P$ denote the fields to which the groups apply.

            Eq. \ref{UDDU} now becomes\begin{equation}\label{UUDDUU}
            D^{\prime}_{\mu,x}U_{B}(x)U_{P}(x)\psi =U_{B}(x)U_{P}(x)D_{\mu,x}\psi.\end{equation} Here $U_{B}(x)=e^{-ia(x)}$ and $U_{P}(x)=e^{-ib(x)}.$  The conditions of Eq. \ref{ABpAB} are replaced by \begin{equation}\label{ABppAB}\begin{array}{c}A^{\prime}_{\mu}(x)=A_{\mu}(x)\\\\ a_{b}B^{\prime}_{\mu}(x)+a_{p}P^{\prime}_{\mu}(x)= a_{b}B_{\mu}(x)+a_{p}P_{\mu}(x)+\partial_{\mu,x}(a(x)+b(x)).\end{array}\end{equation} The second equation is equivalent to a pair of equations as \begin{equation}\label{BpPp}\begin{array}{c} B^{\prime}_{\mu}(x)=B_{\mu}(x)+\frac{\mbox{$\partial_{\mu,x}a(x)$}}{\mbox{$a_{b}$}} \\\\P^{\prime}_{\mu}(x)=P_{\mu}(x)+ \frac{\mbox{$\partial_{\mu,x}b(x)$}}{\mbox{$a_{p}$}}. \end{array}\end{equation}

            It follows from these conditions that both the $\vec{B}$ and $\vec{P}$ fields have mass $0.$  However they are quite distinct. One reason is that $\vec{B}$ is integrable whereas the electromagnetic field is not.  This is a consequence of the Aharonov Bohm effect \cite{AhBo}.

            These results show that inclusion of number scaling effects in the QED Lagrangian are taken care of by addition of terms for the $\vec{A}$ and $\vec{B}$ fields. The Lagrangian is given by
            \begin{equation}\label{LQED}L_{QED}(x)=\bar{\psi}i\gamma^{\mu}(\partial_{\mu,x}+a_{a} A_{\mu}(x)+ia_{b}B_{\mu} (x)+ia_{p}P_{\mu}(x))\psi-m\bar{\psi}\psi -\frac{1}{4}G_{\mu,\nu}G^{\mu,\nu}.\end{equation} Here $\frac{1}{4}G_{\mu,\nu}G^{\mu,\nu}$  is the Yang Mills term where \begin{equation}\label{Gmunu}G_{\mu,\nu}=\partial_{\mu}P_{\nu} -\partial_{\nu}P_{\mu}.\end{equation}Both the $\vec{B}$ and $\vec{P}$ fields have zero mass.  The $\vec{A}$ field can have any mass, including $0.$

            The great accuracy of the QED Lagrangian in describing experimental results \cite{Odom,Gabrielse} without the $\vec{A}$ or $\vec{B}$ fields present means that the coupling constants, $a_{a}$ and $a_{b},$ must be very small compared to the fine structure constant. Another possibility is that the the values of the $\vec{A}$ and $\vec{B}$ fields are close to zero in all regions of space and time in which experiments that test or depend on QED accuracy have been carried out.

            \subsection{Nonabelian gauge theories}
            For nonabelian gauge theories invariance of terms in Lagrange densities is extended to  include  local $SU(n)$ gauge transformations. The covariant derivative is obtained in the same way as for the Abelian gauge theories.  The resulting  covariant derivative differs from that of Eq. \ref{DU1NS} by the addition of a term that is the sum over the $n^{2}-1$ generators of the Lie algebra, $su(n).$ One obtains, \begin{equation}\label{DmuxNA}D_{\mu,x}\psi=(\partial_{\mu,x}+a_{a}A_{\mu}(x) +ia_{b}B_{\mu}(x)+ia_{p}P_{\mu}(x)+ia_{t}w_{\mu}^{a}(x)T^{a})\psi.\end{equation}Here $\vec{w}$ is an $n^{2}-1$ component real vector field and the $T^{a}$ are the generators of $su(n).$ For the simplest case with $n=2,$ the $T^{a}$ are the three Pauli operators.

            \subsection{The Higgs mechanism and $\vec{A}$ and $\vec{B}$ fields}\label{HMABF}
            The presence of the $\vec{A}$ and $\vec{B}$ vector fields in the connections raises the question regarding what physical properties these fields  may have. So far one knows that they must make negligible contributions to the Lagrangian for $QED.$  Another possibility is to see if these fields have any relation to the Higgs boson \cite{Higgs}.

            This simplest way to investigate this is to determine the role, if any of $\vec{A}$ and $\vec{B}$ in the symmetry breaking of the Lagrangian for the Mexican Hat potential. Let $\psi$ be a complex valued scalar field as a section over $\mathfrak{CV}^{\cup}$, Eq. \ref{sumcCV}.  For each $x$ in $M$ $\psi(x)$ is a complex number in $\bar{C}^{cg(x)}_{x}$ in the fiber at $x.$

            The  Lagrangian density for this potential is \cite{Cheng} \begin{equation}\label{LMH}L_{MH}(x)=
            D^{\mu}\psi^{*}D_{\mu}\psi+\mu^{2}\psi^{*} \psi-\lambda(\psi^{*} \psi)^{2} \end{equation}Since $\beta(x)$ is a scalar field, $G_{\mu,\nu}=(\partial_{\nu}\partial_{\mu} -\partial_{\mu} \partial_{\nu})\beta(x)=0.$  The field, $\psi$ can be written as\begin{equation} \label{psi12}\psi= \frac{1}{\sqrt{2}}(\psi_{1}+i\psi_{2}).\end{equation} Here $\psi_{1}$ and $\psi_{2}$ are both real.

            The application of the Higgs mechanism follows the description  in \cite{Cheng}.  From the nonzero value of the potential minimum, given by $\psi^{*}\psi=v^{2}/2 =\mu^{2}/2\lambda$, the field, $\psi$ has a nonzero vacuum expectation value, $$|\langle 0|\psi|0\rangle| =v/\sqrt{2}=\mu/\sqrt{2\lambda}.$$

            The rotational symmetry around the Mexican hat potential is broken by a specific choice, $\langle 0|\psi_{1}|0\rangle =v$ and $\langle 0|\psi_{2}|0\rangle =0.$ Replacement of $\psi_{1}$ by $\psi_{1}'+v,$  where $\psi_{1}'$ is a shifted field, gives objectionable mixing terms in the Lagrangian.  These can be removed by working in the unitary gauge where $\psi$ is replaced by $\psi^{\prime}$. In this case \begin{equation}\label{psixva}\psi^{\prime}(x) =\frac{1}{\sqrt{2}}(v+\theta(x))=\psi(x)e^{-i\phi(x)/v},\end{equation}and defining a new field $B'_{\mu}(x),$ by \begin{equation}\label{Bpmuxp}B'_{\mu}(x)=B_{\mu}(x)+\frac{1}{a_{b}v} \partial_{\mu}\phi(x).\end{equation}

            The covariant derivative becomes, \begin{equation}\label{Dmupsi2} \begin{array}{l}D_{\mu,x}\psi= (\partial_{\mu,x} +a_{a}A_{\mu}(x) +ia_{b}B_{\mu}(x)) \psi^{\prime}e^{i\phi(x)/v}\\\\\hspace{1cm} =\frac{1}{\sqrt{2}}[\partial_{\mu}\theta(x)+ (a_{a}A_{\mu}(x)+ia_{b}B'_{\mu}(x)) (v+\theta(x))]e^{i\phi(x)/v}.\end{array}\end{equation}

            Use of this in the Mexican hat Lagrangian gives mass terms $(1/2)(a_{a}v)^{2}A_{\mu}A^{\mu}$ and $(1/2)(a_{b}v)^{2}B'_{\mu}B'^{\mu}$ for the $\vec{A}$ and $\vec{B}'$ fields. The massless $\vec{B}$  combines with $\phi(x)$ to becomes a massive vector boson, $\vec{B}'.$ The mass, $\mu^{2}/2$ of the $\theta(x)$ field arises from the mass term, $\mu^{2}(\theta(x))^{2}$ in the potential.

            The $\vec{A}$ field is also massive.   However this is not new in that nothing prevents addition of a mass term to the Lagrangian at the outset.  Also there are mixing  terms, $a_{a}A^{\prime}_{\mu}\partial^{\mu}\theta(x)+a_{a}A^{\prime,\mu}\partial_{\mu}\theta(x),$ in the Lagrangian.  Here $A_{\mu}^{\prime}=A_{\mu}(v+\theta).$

            These results show that the Higgs mechanism can be used to combine the massless phase part, $\vec{B},$ of the connection with  $\phi(x)$ to give a massive vector boson, $\vec{B}'.$  The remaining real field, $\theta(x),$ with mass term, $\mu^{2}\theta(x)^{2},$ is the Higgs field \cite{Bailin}.

            The possible relationship of $\vec{A}$   to  the Higgs field is intriguing.  The fact that $\vec{A}$ is integrable raises the possibility that $\theta$ field can be identified with $\alpha.$ in this case, $\vec{A}(x)=\nabla_{x}\theta(x)$ where $\theta(x)$ is a real scalar field, presumably of spin $0.$ This suggests the possibility that  $\alpha(x)$  is the Higg's field.  In this case it would replace  $\theta(x)$ in the above derivation.  Whether this  possibility is true or not is open at this point.  More work is needed.

            \section{Number scaling and geometry}\label{NSG}
            \subsection{Fiber bundles and connections}
            The fiber bundle structure for geometric quantities is quite similar to that for gauge theories. The vector spaces in geometry are tangent spaces over $M$.  The associated scalars are real numbers. The fiber bundle of Eq. \ref{sumcCV} becomes \begin{equation}\label{sumrRT}\mathfrak{RT}^{\cup} =\sum_{r}\mathfrak{RT}^{r}=\{M\times \bigcup_{r}(\bar{R}^{r}\times\bar{T}^{r}),Q,M\}.\end{equation} The  sum is over all positive real numbers. At each point $x$ of $M$ the fiber at $x$ is given by \begin{equation}\label{QrRT} Q^{-1}(x)=\bigcup_{r} (\bar{R}^{r}_{x} \times\bar{T}^{r}_{x}). \end{equation}Since $M$ is flat $\bar{T}^{r}_{x}$ is a representation, at $x$, of all of $M$ at level $r$ in the bundle.

            The components of $\bar{T}^{r}$ are given by \begin{equation}\label{BTr}\bar{T}^{r}=\{T,\pm_{r},\odot_{r},\cdot_{r},v_{r}\}.\end{equation}The operations, $\odot_{r}$ and $\cdot_{r}$ denote respectively  the scalar product between two vectors and the product of a scalar and a vector. $v_{r}$ denotes an arbitrary vector.  The associated  real number structure, $\bar{R}^{r}$ is given by Eq. \ref{BRs}.  The fiber bundle, $\mathfrak{RT}^{\cup}$, takes account of the fact that, for each scaling factor $r,$  one has a tangent space $\bar{T}^{r}$ and its associated real scalar field, $\bar{R}^{r}.$

            The relative scaling of the real numbers results in relative scaling for the tangent spaces. The relative scaling of $\bar{T}^{t}$ on $\bar{T}^{s}$ is based on that for $\bar{R}^{t}$ and $\bar{R}^{s}.$ One has from Eq. \ref{BRaRCts}\begin{equation}\label{BRts}\bar{R}^{t}_{s}= \{R,\pm_{s},\frac{s}{t}\times_{s}, \frac{t}{s}(-)^{-1_{s}},<_{s},0_{s},\frac{t} {s}1_{s}\}. \end{equation} The components of corresponding tangent space, $\bar{T}^{t}_{s}$, are given by \begin{equation}\label{BTts}\bar{T}^{t}_{s}=\{T,\pm_{s},\frac{s}{t}\odot_{s},\frac{s}{t} \cdot_{s},\frac{t}{s}v_{s}\}.\end{equation}

            The structure group for the fiber of $\mathfrak{RT}^{\cup}$ is denoted by $W_{RT}.$   The group elements, $W_{RT,s}$ for each positive $s$ in $GL(1,R),$ act on the fiber elements according to  \begin{equation}\label{WsRT}W_{RT,s} \bar{R}^{r}\times\bar{T}^{r}=\bar{R}^{sr}\times \bar{T}^{sr}. \end{equation}

            The properties of connections on $\mathfrak{RT}^{\cup}$ are quite similar to those for $\mathfrak{CV}^{\cup}.$ Here connections are maps of  components of fibers at point $x+\vec{dx}$ to $x.$  The maps have two components, a horizontal one that maps $\bar{R}^{t}_{x+\vec{dx}} \times\bar{T}^{t}_{x+\vec{dx}}$ to $\bar{R}^{t}_{x}\times \bar{T}^{t}_{x}$ and a vertical one that maps $\bar{R}^{t}_{x}\times\bar{T}^{t}_{x}$ to $\bar{R}^{t}_{s,x}\times\bar{T}^{t}_{s,x}.$ Here $t$ and $s$ are both real scaling values.

            For the bundle $\mathfrak{RT}^{\cup}$ the vertical component of connections can be implemented by  a real valued smooth function  $f$ with domain $M.$  Here the function $f$ will be taken to be the modulus of the function, $g$, used for derivatives in gauge theories.  From Eq. \ref{gxe} one has that \begin{equation}\label{feax}f(x)=e^{\alpha(x)}.\end{equation}

            \subsection{Scaling of geometric quantities}\label{SGQ}

            As noted in the introduction, the scaling introduced here is different from conformal scaling \cite{Ginsparg,Gaberdiel} in that angles between vectors as well as vector lengths are scaled.  For example, let $\phi_{t}$ be the  angle, as a real number value,  between two vectors in $\bar{T}^{t}$,  The angle value, $\phi_{s}$ between the same two vectors in $\bar{T}^{s}$ is the same number value in  $\bar{R}^{s}$ as $\phi_{t}$ is in $\bar{R}^{t}.$  However $\phi_{t}$ corresponds to the scaled value, $(t/s)\phi_{s},$ in $\bar{R}^{t}_{s}.$

            This scaling does not affect trigonometric relations in that all relations are preserved under scaling.  These are examples of the more general property of scaling in that all equations involving numerical values are preserved under scaling.

            As a specific example consider the relation, $\sin^{2}(\phi)+\cos^{2}(\phi)=1.$ Since both $\sin(\phi)$ and $\cos(\phi)$ are analytic functions of the angles, it follows that $\sin^{t}(\phi_{t})$ corresponds to $\sin^{t}_{s}(t\phi_{s}/s)=(t/s)\sin^{s}(\phi_{s})$ in $\bar{R}^{t}_{s}\times\bar{T}^{t}_{s}.$  A similar relation holds for $\cos(\phi).$ The scaling of multiplication in $\bar{R}^{t}_{s}$ results in the following equivalences $$\sin^{t}(\phi_{t})^{2} +_{t}\cos^{t}(\phi)^{2}=1_{t}\leftrightarrow \frac{t}{s}\sin^{s}(\phi_{s})^{2}+_{s}\frac{t}{s} \cos^{s}(\phi_{s})^{2}=\frac{t}{s}1_{s}\leftrightarrow\sin^{s}(\phi_{s})^{2}+_{s} \cos^{s}(\phi_{s})^{2}=1_{s}.$$

             The reason trigonometric relations like the one just described are unaffected by scaling is that $\sin(\phi)$ and $\cos(\phi)$  are local quantities.  This is seen in the fiber bundle $\mathfrak{RT}^{\cup}$ by the fact that these two quantities are number values  in a fiber at an arbitrary but fixed location $x.$  They are not localized versions of nonlocal quantities. such as the derivatives in gauge theories.\footnote{Derivatives are referred to here as nonlocal even though they are local limits of nonlocal quantities.}

             Scaling factors are present in an essential way in the description of nonlocal geometric quantities.   These include integrals and derivatives of functionals over $M$. Two examples are considered here.  They are path lengths and geodesic equations.

             \subsubsection{Path lengths}
              Path lengths are simple  examples of nonlocal geometric properties. As such they show the effects of number scaling. The usual expression for the length of  a timelike  path, $p,$ from $x$ to $y$, parameterized by a real variable, $\gamma,$  is given by\begin{equation}\label{Lp}\begin{array} {c}L(p)=\int \sqrt{-\eta_{\mu,\nu}\frac{dp^{\mu}} {d\gamma}\frac{dp^{\nu}}{d\gamma}}d\gamma.\end{array} \end{equation} Here $\eta_{\mu,\nu}=-1,1,1,1$ is the metric tensor for Minkowski space.

             The tangent bundle viewpoint taken here results in two descriptions of the path length in Eq. \ref{Lp}.  One is local and the other is  nonlocal.  The local description regards the integral as being taken over a tangent space $\bar{T}^{t}_{x}$  at level $t$ in the fiber at $x$. A coordinate chart, $\phi^{t}_{x}$ that maps $M$ to a coordinate system,  $(\mathbb{R}^{t})^{4}_{x}$ in $\bar{T}^{t}_{x}$ is used to lift the path, $p$ to a path in $\bar{T}^{t}_{x}.$ The  path parameters $\gamma$ become real parameters in $\bar{R}^{t}_{x}.$

             The local lengths, $L_{t,x}(p),$ as number values of the length all have the same value in their respective number structure, $\bar{R}^{t}_{x}.$ However $L_{t,x}(p)$ corresponds to the value $(t/s)L_{s,x}(p)$ in $\bar{R}^{s}_{x}.$  This is the value, in $\bar{R}^{s}_{x},$  of the number in $R$ that has value $L_{t,x}(p)$ in $\bar{R}^{t}_{x}.$  This shows that the only effect of scaling on local path lengths is multiplication of the length by a relative scale factor.

             The nonlocal description of the path length is different in that each value of the square root  integrand is in a fiber at a different location.   For each parameter value $\gamma$  the square root integrand  at level $t$ is a value in $\bar{R}^{t}_{p(\gamma)}.$ The derivative, $dp/d\gamma$ is defined for the location, $p^{t}(\gamma_{t}),$  for the lifted path, $p^{t}$ in $\bar{T}^{t}_{p(\gamma)}$  for parameter $\gamma_{t}.$ This is the same value in $\bar{R}^{t}_{p(\gamma)}$ as $\gamma$ is in the global real number structure, $\bar{R}.$

             For this setup the integral of Eq. \ref{Lp} is not defined.   This is fixed by use of connections to map the integrands from $p(\gamma)$ to a common location, $x$.  If no scaling is present, ($f$ is a constant), then only the horizontal component of the connections is active. This component  maps  the square root integrand as a number value in $\bar{R}^{t}_{p(\gamma)}$  at fiber location $p(\gamma)$ to the same value in $\bar{R}^{t}_{x}$ at fiber location $x.$

             The resulting expression for the length is given by \begin{equation}\label{Lpx}L(p)^{t}_{x}=\int_{x} \sqrt{-\eta_{\mu,\nu}\frac{dp^{\mu}} {d\gamma}\frac{dp^{\nu}}{d\gamma}}d\gamma_{t}\end{equation}  This is the same value as is the  local expression for the length at level $t$ in the fiber at $x.$.  It shows that  parallel transformations by purely horizontal connections of nonlocal expressions provide nothing new over the local length expressions.

             The situation changes if vertical components are present in the connections.  In  this case scaling factors are included with the integrand.  For scaling described by the function $f,$ the length of path $p$ at reference point $x$ becomes, \begin{equation}\label{Lfxp}L(p)^{tf(x)}_{x}=\int \frac{f(p(\gamma))}{f(x)}\sqrt{-\eta_{\mu,\nu}\frac{dp^{\mu}} {d\gamma}\frac{dp^{\nu}} {d\gamma}}d\gamma_{tf(x)}=\int e^{\alpha(p(\gamma))-\alpha(x)}\sqrt{-\eta_{\mu,\nu}\frac{dp^{\mu}} {d\gamma}\frac{dp^{\nu}} {d\gamma}}d\gamma.\end{equation} This is an integral of path length values in $\bar{R}^{tf(x)}_{x}.$   The factor $f(x)^{-1}$ can be moved outside the integral as it is a constant. Note that the case of no scaling is the special case where $f$ is a constant over all of $M$.

            Eq. \ref{Lfxp} holds for time like paths.  For space like paths, $-\eta_{\mu,\nu}$ is replaced by $\eta_{\mu,\nu}.$ For null paths, such as those travelled by light rays, the path length is $0.$  This is the only case for which the path length is independent of scaling. It is a consequence of the fact that $0$ is the only number whose value is unaffected by scaling.

\begin{figure}[h!]\begin{center}\vspace{2cm}
            \rotatebox{270}{\resizebox{170pt}{170pt}{\includegraphics[170pt,200pt]
            [520pt,550pt]{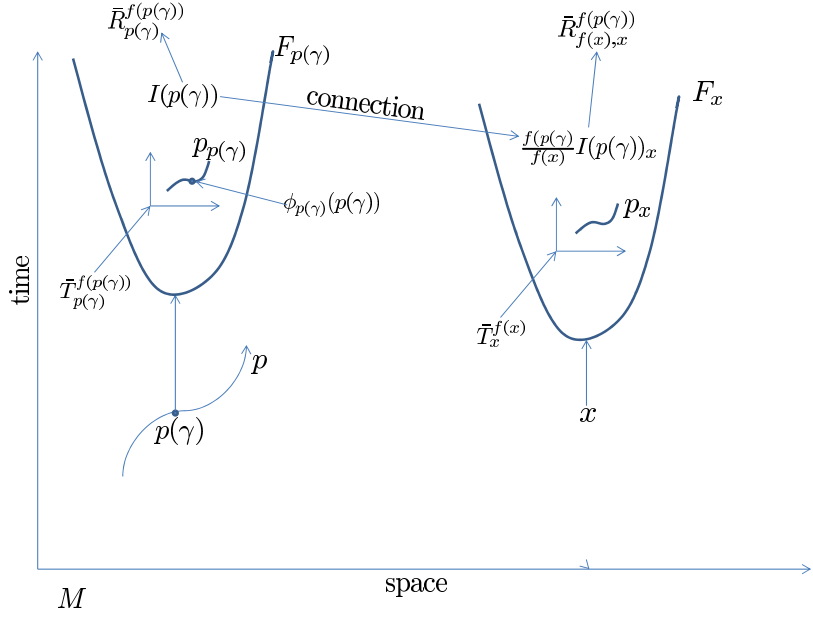}}}\end{center}\caption{Representation of some contents of  fibers $F_{p(\gamma)}$ and $F_{x}$ at point $p(\gamma)$ and a reference point, $x$ on $M$. Space dimensions of $M$ are reduced to one for presentation ease. The tangent spaces at levels $f(p(\gamma ))$ and $f(x)$ with path representations, $p_{p(\gamma)}$ and $p_{x}$, in the tangent spaces are shown. As a number value in $\bar{R}^{f(p(\gamma))}_{p(\gamma)}$, $I(p(\gamma))$  is the integrand value for the point, $\phi_{p(\gamma)}(p(\gamma))$ in the lifted path that corresponds to the point, $p(\gamma)$ in $M$.  Connections map the single integrand values in the fibers at each point, $p(\gamma)$ to scaled values in the fiber at the reference location, $x.$  }\label{FBa}\end{figure}

            Figure \ref{FBa} shows a schematic representation of the nonlocal representation of path lengths.  For each value of $\gamma$ the fiber at  $p(\gamma)$ contains  a tangent space $\bar{T}^{f(p(\gamma)}_{p(\gamma)},$ a  coordinate point $\phi_{p(\gamma)}(p(\gamma)),$ and a value, $I(p(\gamma)),$ of the length integrand in $\bar{R}^{f(p(\gamma))}_{p(\gamma)}.$ Just one of these fibers is shown.  The fiber at the reference location, $x$ is shown along with the vertical and horizontal components of the connection map of the integrand in the fiber at $p(\gamma)$ to the corresponding integrand in the fiber at $x.$

            Eqs. \ref{Lpx} and \ref{Lfxp} show that the length of space and time like paths are affected by scaling. One consequence of this is that the proper time, measured by clocks carried along a time like path, is affected by scaling. This follows from replacement  of the arbitrary parameter $\gamma$ by the proper time, $\tau,$ and  the use of \cite{Carroll} \begin{equation}\label{dtau}d\tau=(-\eta_{\mu,\nu}dp^{\mu} dp^{\nu}/d\gamma d\gamma)^{1/2}d\gamma.\end{equation}

            At reference point, $x,$ of $M,$ the elapsed proper time along $p$ from $0$ to $\mathcal{T}$ is given by \begin{equation}\label{Tau}\mathcal{T}_{\alpha}=e^{-\alpha(x)}\int_{0}^{\mathcal{T}} e^{\alpha(p(\tau))}d\tau.\end{equation} Here $\mathcal{T}_{\alpha}$ is the elapsed proper time in the presence of scaling. In the absence of scaling $\alpha(p(\tau))=\alpha(x)$ for all $\tau.$

            The choice of reference locations for the path lengths in Eqs. \ref{Lfxp} and \ref{Tau} is not limited to $x$ or any point on the path.  Any location, $z,$ in $M$ can be chosen. The resulting path length, as a numerical value in the fiber, $F_{z},$ is obtained by replacing $x$ by $z$ in Eqs. \ref{Lfxp} and \ref{Tau}.

            \subsubsection{Geodesics}

            Geodesics are also affected by number scaling. The minimum path length between $x$ and $y$ is found by varying $L(p)^{tf(x)}_{x}$ with respect to $p$.   Addition of a small path change, $\delta{p},$ to $p$  gives \begin{equation}\label{Lpxrd}\begin{array}{c} L(p+\delta p)^{tf(x)}_{x}=\int e^{-\alpha(x)+\alpha(p(\gamma)+\delta{p})} \sqrt{-\eta_{\mu,\nu} \frac{d(p^{\mu}+\delta{p}^{\mu})} {d\gamma}\frac{d(p^{\nu}+\delta{p}^{\nu})}{d\gamma}}d\gamma.\end{array} \end{equation}

            The Euler Lagrange equations for $L(p )$ are obtained by expansion of Eq. \ref{Lpxrd} to obtain \begin{equation}\label{Lpxrd1}\begin{array}{c}L(p)_{x}+L(\delta{p})_{x}=e^{-\alpha(x)}\int e^{\alpha(p(\gamma))} [1+\frac{d\alpha(p)}{dp^{\mu}}\delta{p}^{\mu}]\\\\\hspace{.5cm}\times (-\eta_{\mu,\nu}\frac{dp^{\mu}}{d\gamma}\frac{dp^{\nu}}{d\gamma})^{1/2} [1+ (-\eta_{\mu,\nu}\frac{dp^{\mu}} {d\gamma}\frac{dp^{\nu}}{d\gamma})^{-1}(-\eta_{\mu,\nu}\frac{dp^{\mu}}{d\gamma} \frac{d(\delta{p}^{\nu})}{d\gamma})]d\gamma.\end{array} \end{equation}Here $e^{-\alpha(x)}$ has been moved outside the integral as a constant factor, and a Taylor expansion of the exponent has been used. The factor in the second line was obtained by expanding the terms in the square root to first  order, factoring out  $(\eta_{\mu,\nu}dp^{\mu}dp^{\nu}/d\gamma d\gamma)^{1/2},$  and expanding the resulting square root \cite{Carroll}. Removal of the $L(p)$ component and changing the arbitrary variable, $\gamma,$ to the proper time, $\tau$  as in Eq. \ref{dtau}  gives\begin{equation}\label{delLp} \delta{L(p)}=\delta{\tau(p)}=e^{-\alpha(x)}\int e^{\alpha(p(\tau))} (-\eta_{\nu,\mu}\frac{dp^{\nu}}{d\tau} \frac{d(\delta{p}^{\mu})}{d\tau}+\frac{d\alpha(p)} {dp^{\mu}}\delta{p}^{\mu})d\tau=0.\end{equation}

            Integration of the first term by parts and assuming $\delta{p^{\mu}}=0$ at the integral endpoints gives, after index relabeling, \begin{equation} \label{delLp0} \delta{\tau(p)}= e^{-\alpha(x)}\int e^{\alpha(p(\tau))}[\eta_{\nu, \mu}\frac{d}{d\tau}\frac{dp^{\nu}}{d\tau} +\eta_{\nu,\mu}\frac{d}{dp^{\rho}}\alpha(p)\frac{dp^{\rho}} {d\tau}\frac{dp^{\nu}}{d\tau} +\frac{d\alpha(p)}{dp^{\mu}}]\delta{p}^{\mu}d\tau=0. \end{equation} This equation is satisfied if the terms in the square brackets sum to $0.$ Multiplying by $\eta^{\nu,\mu}$  gives with some index relabelling \begin{equation}\label{geod1}\frac{d}{d\tau}\frac{dp^{\nu}} {d\tau}+ \frac{d\alpha(p)}{dp^{\rho}}\frac{dp^{\rho}} {d\tau}\frac{dp^{\nu}}{d\tau} +\eta^{\nu,\rho}\frac{d\alpha(p)} {dp^{\rho}}=0.\end{equation}

            This is the geodesic equation in the presence of number scaling.  Comparison of this with the equation from general relativity, \cite{Carroll} \begin{equation}\label{geodgr}\frac{d^{2}p^{\nu}}{d\tau^{2}}+\Gamma^{\nu}_{\rho,\mu} \frac{dp^{\rho}}{d\tau}\frac{dp^{\mu}}{d\tau}=0,\end{equation} shows that the Christoffel symbol is $0$ unless $\mu =\nu.$ In this case $\Gamma^{\nu}_{\rho,\nu}=d\alpha/dp^{\rho}$, independent of $\nu.$

            One can use the chain rule of differentiation to write Eq. \ref{geod1} in the revealing form, \begin{equation}\label{geodcov}[\frac{d}{d\tau}+\vec{A}(p(\tau))\cdot\nabla_{\tau}p] \frac{dp^{\nu}}{d\tau}+\eta^{\nu,\mu}A_{\mu}(p(\tau))=0. \end{equation}The term in the square brackets is similar to a covariant derivative.  It includes some of the effects of number scaling. The remaining effects are shown by the presence of the term, $\eta^{\nu,\mu} A_{\mu}(p(\tau)).$ This arises from the effect of path variation  on the  field $\alpha$ in the exponential, $e^{\alpha(p+\delta{p})}.$

            It is tempting to follow the argument in general relativity \cite{Carroll} by saying that the first two terms of either Eq. \ref{geod1} or \ref{geodcov}  represent the motion of a  particle in the presence of the field, $\alpha.$ In this case the third term acts like a force on the particle, also generated by $\alpha.$ However, one must keep in mind that the effect of $\alpha$, or, more accurately, of the gradient vector field, $\vec{A}(x)=\nabla_{x}\alpha$ on particle motion, arises from the connection between the fibers of the tangent bundle.  This in turn arises from the observation that numbers can be scaled and the scaling factor can depend on space and time.  Note that if $\vec{A}(x)=0$ everywhere, then there is no scaling and Eqs. \ref{geod1} or \ref{geodcov}  describe a straight line path.

            \section{Experimental restrictions on  $\vec{A}$ and $\vec{B}$}\label{ERAB}

            There is no direct physical evidence for the presence of the $\vec{A}$ or $\vec{B}$ fields. This was noted in particular for the QED Lagrangian. The  lack of evidence is also the case for the $\vec{A}$ field in the geometric quantities described here.\footnote{If one used complex geometry, as in \cite{Manin}, then $\vec{B}$ would also affect geometric quantities.} This lack of physical evidence must be reconciled with the properties of the $\vec{A}$ and $\vec{B}$ fields.

            One way to achieve reconciliation is to use a scenario in which $M$ represents cosmological space and time. Time varies from the big bang, about $14\times 10^{9}$ years ago, to the present and beyond. Space positions vary over the observable universe.   General relativity is neglected here.

            This description of $M$ means that the fiber bundles,  described in earlier sections, are bundles of fibers over all of space and time, from the big bang to the present, and over all locations of the universe. It follows that numerical values of geometric and physical quantities are local in that they are expressed as real numerical quantities  in a fiber $F_{x}$ at some location $x.$ Examples of this are  curve lengths in Eq. \ref{Lfxp} and the geodesic equation, Eq. \ref{geodcov}. Both are numerical values in $F_{x}.$

             A direct comparison of these  scaled quantities with experimental results must take account of the fact that all experiments and measurements are necessarily local.  They must occur in a region of space and time  that is either occupied or can be occupied by  observers. The reason is that observers are needed to implement the relevant measurement  procedures.  This is the case whether one is measuring properties of locally produced systems or of incoming radiation emitted by systems at cosmological distances.

             Consider  all geometric or physical quantities whose description is limited to a region of space and time that is or can be occupied by observers.   This includes such things as lengths of paths in the  region or wave packets whose extent is limited to the region.  The lack of evidence for number scaling effects in this region means that both $\vec{A}(x)\approx 0$ and $\vec{B}(x)\approx 0$ to the limits of experimental accuracy in the region.  It says nothing about the values of $\vec{A}(x)$ or $\vec{B}(x)$ outside the region. They can be quite large and vary rapidly in very far away regions in space and time, including those near and after the big bang.

             The  size of this region is not known.   However it must include, at least, regions already occupied by observers. This includes the surface of the earth. Since we, as observers, are in principle capable of interplanetary travel, the region should include the solar system.
             The region should also include  other planets and areas around other nearby stars which are in reasonable communication distances from us. A generous estimate of the size of this region is that of a sphere about $2,500$ light years in diameter \cite{AMSci} roughly centered on the solar system.

             The exact size of this region is not important. What is relevant is that the region is a small fraction of the observable universe. The size  is predicated on the idea that observers in locations outside the region are too far away for us to know that they exist or for us to communicate with them, now or in the future.

            \section{Conclusion}\label{C}

            This work represents some advances over earlier work on the topic of number scaling \cite{BenNOVA}.  These include advances in the explanation of number scaling and in the use of fiber bundles. The advances in number scaling consist of a clearer separation of numbers, as elements of the base set of a type (real, complex, etc.) of number structure,  from the values they have in a particular  scaled structure.

            The essential point to remember is that for each type of number, the elements of a base set, as numbers have no predetermined value.  They acquire values only inside a structure  that satisfies a set of axioms for the type of number being considered.  It was seen that for each number type many different structures can be described where the value of  each of the base set elements depends on the structure containing them. This is the meaning of number scaling.

            An advantage of the use of fiber bundles is that the bundles make a clear distinction between local mathematical structures, as elements of fibers at different locations on a space time manifold, $M,$ and global mathematical structures that do not belong to any fiber.  These exist outside of space and time \cite{Tegmark} and are outside of any fiber bundle whose base space is $M.$  Elements of some global structures serve as connections between the fibers.

            There is much more work to be done.  This includes extension to include the effects of scaling on many other physical  and geometric quantities, besides the few described here, and  expansion to include general relativity. Also more work is needed to determine the physical nature, if any, of the $\vec{A}$ and $\vec{B}$ fields.  One may hope that these fields correspond to  physical fields, but much more work is needed to see if this is true or not.

            Another area that needs additional investigation is the nature of the base set elements of number structures and of other types of structures.  In particular the relationship between these elements and physical systems needs investigation. Quantum mechanical representations of numbers may be useful here.  This is especially the case when one considers that expressions of numbers and number properties, as alphabet symbol strings, are usually described classically.  This follows from the observation that symbols in a string are assumed to have particular intrinsic  values irrespective of whether  or not they are bing examined.

            One point about the effect of number scaling mentioned in earlier work  \cite{BenNOVA} deserves to be emphasized again.  This is that scaling plays no role in comparisons between outputs of experiments with one another or with outputs of computations as theoretical predictions of experimental results.

            This might seem surprising in that  comparison of a  computational output as a number value at $x$ with an  experimental result as a number value at $y$ can only be done locally at some point of $M$. In the presence of scaling, transferral of the number value at $y$ to $x$ requires multiplication of the experimental result by a scaling factor. The scaled number value is then compared locally with the  number value of the computation output.

            The problem is that physics gives no hint of any such scaling. This is not a problem when one realizes that outputs of experiments and computations, as physical processes, are never number values.  They are physical systems in physical states. They can be interpreted as  number values but these are local interpretations that refer to the output locations.

            Comparison of the outputs requires physical transportation of the information in the output states to a common location. At this location, the two output states, interpreted as number values at the same point, are compared \emph{locally}.  No number scaling is involved.

            In conclusion one hopes that this work is a contribution to the more general topic of the basic relationship between mathematics and physics.  This is a topic that has interested this author \cite{BenCTPM2}, and many others  \cite{Tegmark,Wigner,Omnes,Plotnitsky}.  Clearly there is much work to be done.

            \section*{Acknowledgement}
            This material is based upon work supported by the U.S. Department of Energy, Office of Science, Office of Nuclear Physics, under contract number DE-AC02-06CH11357.

                        \end{document}